\documentclass[12pt]{article}
\usepackage{graphicx}
\usepackage{subfig}

\usepackage{amsmath}
\usepackage{amsfonts}
\usepackage{amssymb}
	
\renewcommand \vec[1]{\boldsymbol{#1}}

\newcommand{\address}[1]{\begin{description}
   														\item[]\rm\raggedright #1
   												 \end{description}
   												}
\newcommand \ead[1]{#1}
\newcommand \bsigma {\vec{\sigma}}
\newcommand \imi{i}
\newcommand \enm{e}
\newcommand \dsgn{d}
\newcommand \rmd{\dsgn}
\newcommand \rme{\enm}
\newcommand \rmi{\imi}
\newcommand \fref[1]{fig.\ref{#1}}
\newcommand \Fref[1]{Fig.\ref{#1}}
\newcommand \eref[1]{ \eqref{#1} }
\newcommand {\fl} {}
\newcommand {\ack} {\section*{Acknowledgments}}

\newcommand \R {\mathbb R}

\newcommand{\intd}   {\int\limits_{\R}   \rmd  }
\newcommand{\intdd}  {\int\limits_{\R^2} \rmd^2 }

\newcommand{\trans}{\top}

\begin{document}
  \title{Representations of solutions of the wave equation 
	 based on relativistic wavelets}

\author{Maria Perel$^{1,2}$ and Evgeny Gorodnitskiy$^1$}

\maketitle

\address{$^1$ Department of Mathematical Physics, Physics Faculty, St.Petersburg University,
		Ulyanovskaya 1-1, Petrodvorets, St.Petersburg, 198904, Russia}
	\address{$^2$ Ioffe Physical-Technical Institute of the Russian Academy of Sciences\\
	26 Polytekhnicheskaya, St Petersburg 194021, Russia}
	\ead{perel@mph.phys.spbu.ru}
	
\begin{abstract}
A  representation of solutions of the wave equation with two spatial coordinates in terms of
localized elementary ones is presented. Elementary solutions are constructed from four solutions 
with the help of  transformations of the affine Poincar\'e group, i.e., with the help of 
translations, dilations in space
and time and Lorentz transformations. The representation can be interpreted in terms of the
initial-boundary value problem for the wave equation in a half-plane. It gives the solution as an
integral representation of two types of solutions: propagating localized  solutions
running away from the boundary under different angles and packet-like surface waves running along
the boundary and exponentially decreasing away from the boundary. Properties of elementary
solutions are discussed. A numerical investigation of coefficients of the decomposition is carried
out. An example of the field created by sources moving along a line with different speeds is
considered, and the dependence of  coefficients on  speeds of sources is discussed.
\end{abstract}

\section{Introduction}
 The construction of exact representations of solutions of the wave equation in terms of elementary
localized solutions   is the subject of the present paper.
Fourier analysis yields a representation of solutions in terms of plane waves. An analogue of
Fourier analysis in our considerations is  continuous wavelet analysis.
We use here a special version of  continuous wavelet analysis based on the transformations of
the affine Poincar{\'e} group  \cite{Antoine} (Wavelets based on  the Poincar\'e group was 
discussed also in \cite{Klauder-Streater}, \cite{Bohnke-1991}.) 
The representation obtained by us is associated with  the initial-boundary value problem in the
half-plane. It is given in terms of the affine Poincar{\'e} wavelet transform of time-dependent 
boundary
data on a line, which is the boundary of the half-plane. The  wavelet transform is efficient in
processing of functions of time and coordinate,  which are  results of observations and
which have a multiscale structure \cite{Antoine}. The result of propagation of  data is given by 
us in terms appropriate for data processing.

The representations of solutions of the non-stationary wave equation  as superpositions of
elementary localized solutions have been presented for the first time in the work of Kaiser
\cite{Kaiser-1994}. They were based on the analytic wavelet transform, and he
used  very special spherically symmetric elementary solutions. Representations of  solutions  in 
terms of elementary solutions from a wide class were proposed in \cite{Perel-Sidirenko-2003}, 
\cite{Perel-Sidorenko-2006}, \cite{Per-Sid-JPA2} and discussed in \cite{Per-Sid-Gor2}.
That representation was based on continuous wavelet analysis. Elementary solutions are
constructed from two chosen ones by means of a similitude group of transformations that contains
shifts, scalings, and rotations. In present  work, a
solution of the non-stationary boundary problem for the wave equation has been decomposed  in 
localized elementary solutions by means of the space-time version of wavelet analysis. Elementary 
solutions are constructed from four chosen ones by means of an affine
Poincar\'e group of transformations that contains shifts and scaling both in space and in time and 
the Lorentz transformations. Preliminary results were reported in \cite{Perel-2009}. The affine 
Poincar\'e wavelet transform (APWT) is a
coefficient in the decomposition of solutions. The region of integration is determined by the 
values of parameters under which the APWT is above a certain level.
Numerical implementation and examples of calculation of APWT were discussed in our work
\cite{Gorodnitskiy-Perel-2011}.

For the sake of simplicity we consider here only the case of two spacial dimensions.

In section \ref{sec:CWT},
we  recall  affine Poincar\'e wavelet transform theory. In section \ref{sec:solutions}, we develop
Poincar\'e wavelet theory for solutions and construct representations of  solutions of the wave
equation based on it. Then we give examples of mother solutions and discuss properties of
elementary solutions obtained from them.   The implementation and
interpretation of Poincar\'e wavelet transform   are considered in section \ref{sec:examples} with 
several examples.

We suggest to determine mother wavelets by means of their Fourier integrals which has not explicit 
form in the coordinate domain. However we may take as mother solutions  exact particle-like 
solutions found in \cite{Kiselev-Perel-1999}, \cite{Kiselev-Perel}, \cite{Perel-Fial}, 
\cite{Per-Sid-JPA1} which are given by simple explicit formulas both in the coordinate and in the 
Fourier domains.  The review of another exact localized solutions is done in \cite{Kiselev}, 
\cite{Recami}. Solutions obtained by Lorentz transformations are discussed in \cite{Saari-2004}.

A representation obtained of solutions can find applications in the non-stationary diffraction 
theory, in the propagation and scattering of multiscale fields, for example, in seismology. We 
hope that our results may be useful also in astronomy to determine velocities of relativistic 
objects or to deal with time-dependent problems of scattering. 

\section{Basics of affine Poincar\'e wavelet analysis}\label{sec:CWT}
First steps in the application of  wavelet analysis methods  are the definition of a
class of functions ${\cal H}$ considered, a choice of a mother wavelet $\psi \in {\cal H},$ and 
the construction
of a family of wavelets. The mother wavelet is  a fixed function. A  family of wavelets is
functions obtained from the mother wavelet by means of transformations of a group. The set of 
wavelets should be dense in the chosen space ${\cal H}$. Then
it should be proved   that  any function $f \in {\cal H}$ can be represented as an integral
superposition of wavelets.

It is useful to consider  together functions and their Fourier transforms.
Below we introduce the notation
\begin{equation}
\vec{\chi}=\binom{ct}{x}, \quad \bsigma=\binom{\omega/c}{k_x},
\end{equation}
and define the Minkovsky inner product as follows:  $(\vec\sigma, \vec\chi)= -\omega t + k_x x$.
The method of
wavelet analysis itself is not associated with the wave equation. We apply it to the boundary data
function $f(\vec{\chi}) \equiv f(ct, x)$ dependent only on one spatial coordinate $x$ and time
$t.$ \footnote{Here $c$ is a parameter of the dimension of speed. We use the notation, taking in
mind further applications to the wave equation in two spatial dimensions where the wave number is
$k=|\vec k|,$  the frequency is $\omega = \pm c k$ and $c$ is the speed of light.}
The Fourier transform of any $f(\vec{\chi}) \in \mathbb{L}_2$  reads
\begin{equation}
\hat{f}(\vec\sigma)=\intdd \vec\chi \, f(\vec\chi)\rme^{-\rmi (\vec\sigma, \vec\chi)}.
\end{equation}

 Lorentz transformations enables one to connect the
coordinate $x$ and the time $t$ in a stationary coordinate system with
the coordinate $x'$ and the time $t'$ in a moving one with  speed $v$. It reads
\begin{eqnarray}\label{eq:lorentz}
   \binom{ct'}{x'} & = \Lambda_\phi \binom{ct}{x},
   \\
   \Lambda_\phi & =\left(\begin{array}{cc}
   								 					\cosh\phi & -\sinh\phi
   									 			  \\
  													-\sinh\phi & \cosh\phi	
  									  	 \end{array}
  							  \right)	
    						 ,\tanh \phi = \frac{v}{c},
\end{eqnarray}
where  $\phi$ is called the rapidity and $v$ is the speed of the moving frame. The frequency and 
the
wave number in the moving frame and in the stationary frame given by  vectors $\bsigma'$
and $\bsigma,$ respectively, are related as
$\bsigma' = \Lambda_{-\phi} \bsigma.$

We are going to clarify the choice of the space ${\cal H}.$ We choose any $\psi(ct,x) \in
\mathbb{L}_2$.  Construct a family of wavelets
with transformations of the affine Poincar\'e group, i.e., with shifting, scaling and the Lorentz
transformations:
\begin{eqnarray}\label{eq:wavelet_family}
&\psi_\mu(\vec\chi)=\frac{1}{a}\psi\left(\Lambda_{-\phi} \frac{\vec{\chi} - \vec{b}}{a}\right) \, ,
 \vec{b}=\binom{c\tau}{b_x}\,, \\
&\mu=\{ \vec{b}, a, \phi \}. \,
\label{eq:mu_definition}
\end{eqnarray}
A wavelet $\psi_\mu(\vec\chi)$  for $a=1,$ $\vec b=0$  represents a wave packet in the
stationary frame, such that it is the mother wavelet in the moving frame. The constructed family 
is not
dense in the whole $\mathbb{L}_2.$
To show this,   we analyze the Fourier transform  of wavelets
\begin{equation}\label{eq:wav-Fourier}
\widehat{\psi_\mu}(\vec\sigma) = a\widehat{\psi}(a \Lambda_{-\phi} \vec{\sigma})
\rme^{-\rmi (\vec{\sigma}, \vec{b})}\,
\end{equation}
and show that we cannot go beyond the region
bounded by the lines $\omega = \pm c k_{x}$ under the action of Lorentz transformations. For
example, a point from the domain $k>0, |k_x|<k$
cannot be mapped to any point in the domain
 $k_x>0, |k|<k_x,$ by choosing $a$ and $\phi$.

The explanation of this fact is below.
 Let the support of the Fourier transform of the mother wavelet in the moving frame be
concentrated near a point $\vec\sigma_0',$    i.e., $\hat{\psi}(\vec\sigma')\equiv 0$  outside a 
neighborhood of $\vec\sigma_0'.$ The support of this packet in the stationary frame lies in the
neighborhood of $\vec\sigma_0,$ such that $\vec\sigma_0 = a^{-1} \Lambda_{\phi} \vec\sigma'_0.$  
It is
easy to check that the point $(\omega_0/c, k_{x0})$ lies on the hyperbola $(\omega_0/c)^2 -
k_{x0}^2 = a^{-2} ( (\omega_0'/c)^2 - k_{x0}^{'2} ).$ For any prescribed $(\omega_0, k_{x0})$ we
can find  a unique parameter $a>0.$ The parameter $\phi$ can be found by the relation $\tanh(\phi
+ \phi') = c k_{x0}/\omega_0,$ where $\tanh(\phi') = c k'_x/\omega'.$
 As $\phi$ increases, the point $(\omega_0/c, k_{x0})$ tends to the asymptotes of the hyperbola
 $\omega = \pm c k_{x}$.  Let $\omega_0'>0,$ $c|k_{x0}'| < \omega_0'$. Then for any $a$ and $\phi$
 the point $(\omega_0/c, k_{x0})$ lies in the same region: $\omega_0>0,$ $c |k_{x0}| < \omega_0$.

\begin{figure}[htbp]
\subfloat[Domains ${\cal{D}}_1$ and ${\cal{D}}_2$]
{\includegraphics[keepaspectratio,width=0.49\textwidth]
{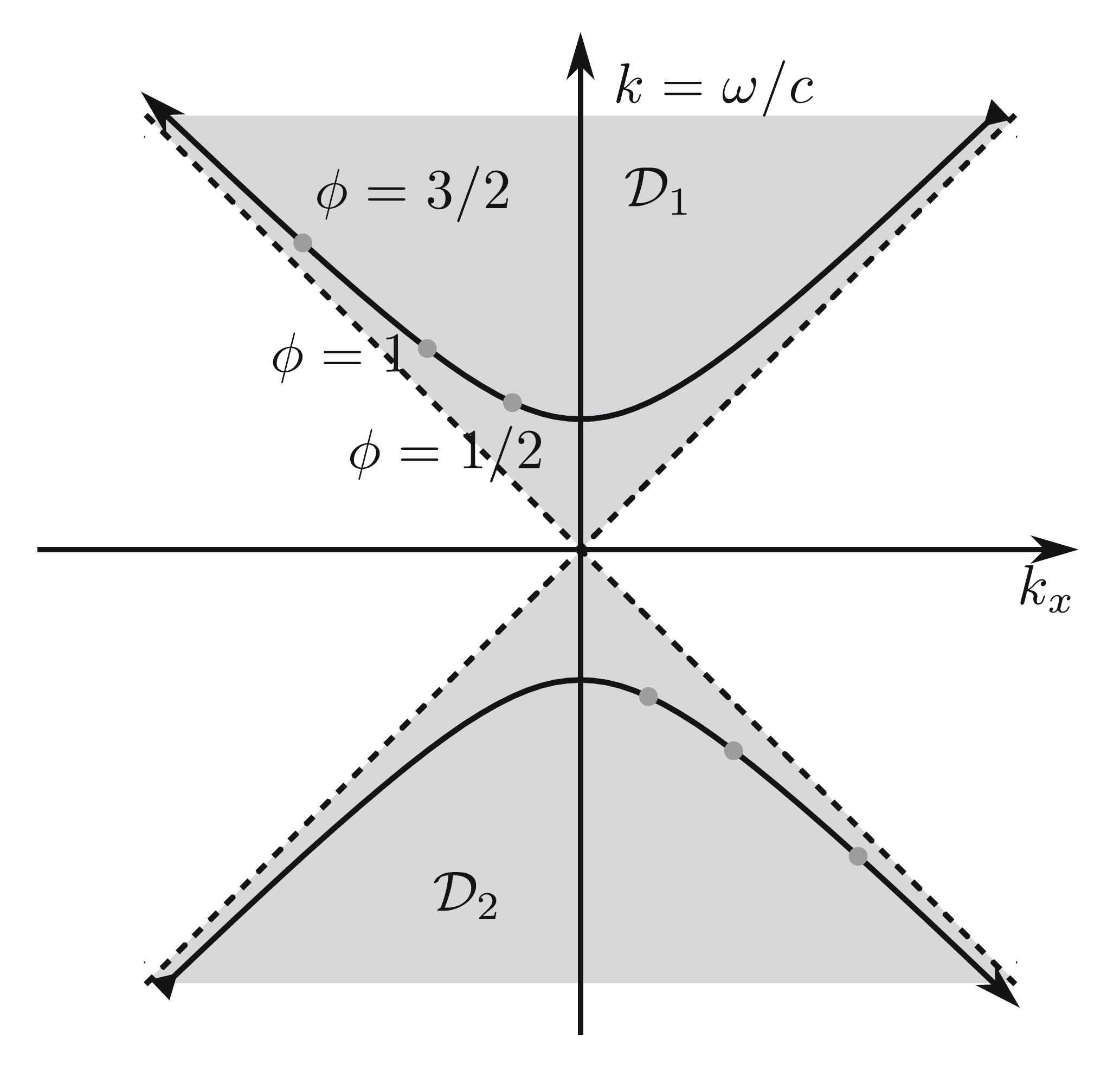}\label{fig:domains:d12}}
\subfloat[Domains ${\cal{D}}_3$ and ${\cal{D}}_4$]
{\includegraphics[keepaspectratio,width=0.49\textwidth]
{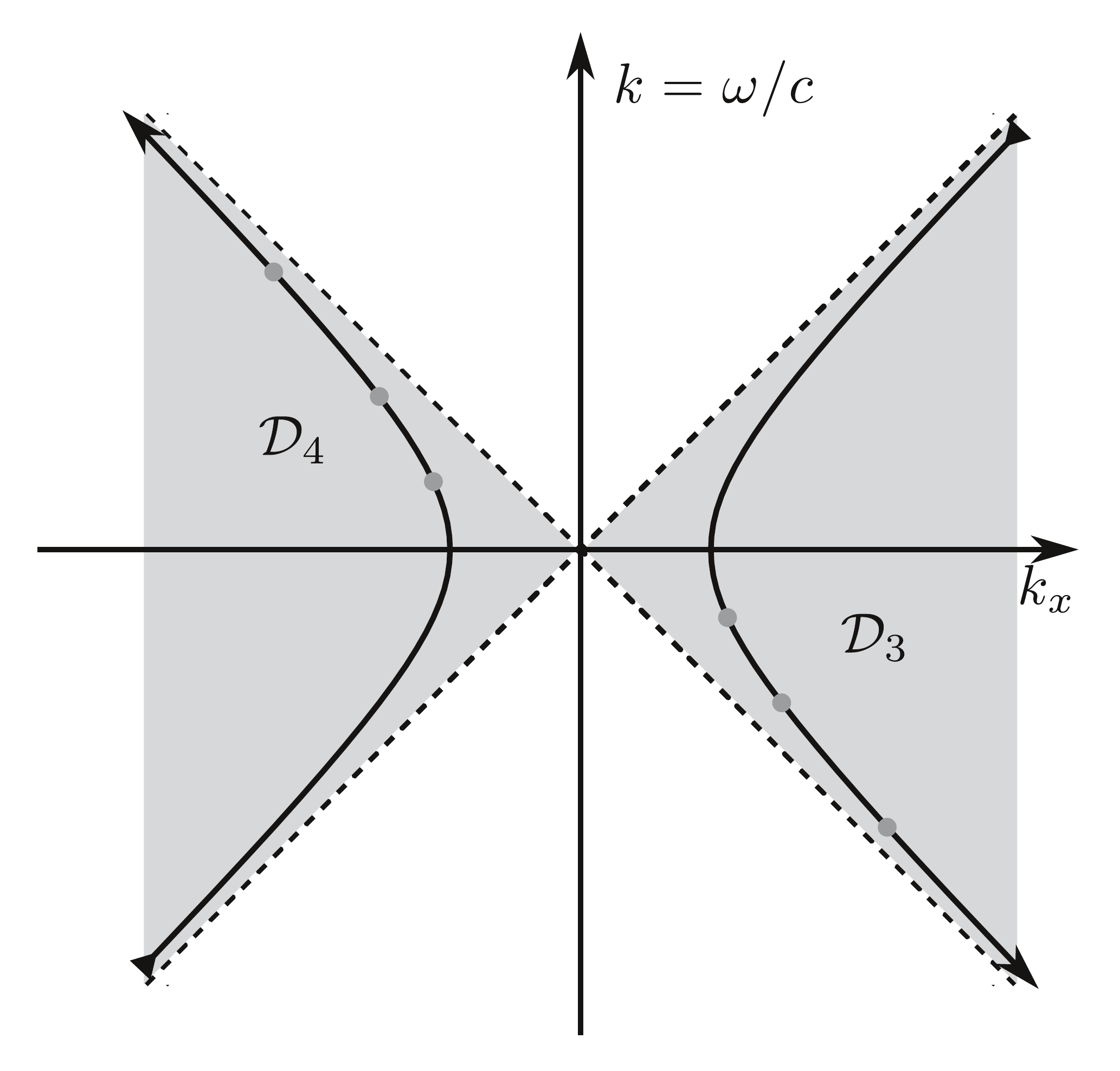}\label{fig:domains:d34}}
\caption{Domains ${\cal{D}}_j$ that are  supports of the Fourier transform of
functions from ${\cal{H}}_j$, $j=1,2,3,4$. The set of points obtained by Lorentz transformations
from the point with $\phi=0$ is a hyperbola. Arrows on
hyperbolas show the direction of point movement from $\phi=-\infty$ to $\phi=+\infty.$}
\label{fig:domains}
\end{figure}

 Therefore we should decompose the space ${\mathbb{L}}_2(\R^2)$  into a direct sum:
 ${\mathbb{L}}_2(\R^2)=\bigoplus\limits_{j=1}^4 {\cal{H}}_j$ taking into account their Fourier
 transforms. The space ${\cal{H}}_j$ comprises functions $f_j \in {\mathbb{L}}_2(\R^2),$ which have
 the support of their Fourier transform in  ${\cal{D}}_j$ (see \fref{fig:domains}) :
\begin{equation}\label{eq:L_domains}
f_j(\vec{\chi})=\frac{1}{(2\pi)^2}\int\limits_{{\cal{D}}_j} \rmd^2\vec{\sigma} \,
\hat{f}(\vec{\sigma})
 \rme^{\rmi(\vec{\sigma}, \vec{\chi})} \, .
\end{equation}
The mother wavelet is any function  $\psi_j(\vec\chi)\in{\cal{H}}_j$ that  satisfies an
admissibility condition:
\begin{equation}\label{eq:admissibility_cond}
C_j= \int\limits_{{\cal{D}}_j} \rmd^2 \vec{\sigma} \,
\frac{|\hat{\psi}_j(\vec\sigma)|^2}{|(\omega/c)^2 - k^2_x|} < \infty.
\end{equation}

The family of wavelets is constructed by formulas (\ref{eq:wavelet_family}).

The affine Poincar\'e wavelet transform $F_j(\mu)$ of the function $f_j$ is defined as follows 
\cite{Antoine}:
\begin{equation}\label{eq:Poincare_CWT}
F_j(\mu)=
\intdd \vec{\chi} \, f_j(\vec\chi) \overline{\psi_{j\mu}(\vec{\chi})},
\quad \mu=\{ \vec{b}, a, \phi \}.
\end{equation}
We denote functions from $\mathbb{L}_2$ by Latin lowercase letters and their wavelet transforms by
Latin uppercase letters.
It is not necessary to extract $f_j(\vec\chi)$ from $f(\vec\chi)$ to calculate $F_j(\mu)$, as 
follows from the Plancherel equality:
\begin{equation}\label{eq:Plancherel_fourier}
(f_j, \psi_{j\mu}) = \frac{1}{(2\pi)^2} (\hat{f}_j, \hat{\psi}_{j\mu})= \frac{1}{(2\pi)^2}
(\hat{f}, \hat{\psi}_{j\mu}) = (f, \psi_{j\mu}).
\end{equation}

The main fact that yields the  decomposition of solutions is the Plancherel equality for the 
affine Poincar\'e wavelet
transform \cite{Antoine}:
\begin{equation}\label{eq:Plancherel_eq}
(f_j, g_j)_{\mathbb{L}_2} = \frac{1}{{C_{j}}} \int \rmd\mu \, F_j(\mu)\overline{G_j(\mu)},
\end{equation}
where
\begin{equation}\label{eq:dmu}
\int \rmd\mu \equiv {\intd \phi}
{\int\limits_{0}^{\infty}\frac{\rmd a}{{a^3}}}{\intdd \vec b}.
\end{equation}
Equality  \eref{eq:Plancherel_eq} is valid for any $f_j, g_j \in {\cal H}_j$ if the admissibility
condition
\eqref{eq:admissibility_cond} is fulfilled.

The reconstruction formula following from the Plancherel equality  reads
\begin{equation}\label{eq:poinc-WT_reconst}
f_j(\vec \chi)= \frac{1}{{C_{j}}} \int \rmd \mu \, F_j(\mu)\psi_{j\mu}(\vec{\chi}).
\end{equation}
See the \ref{app:rec} for more detail.
Any $f(\vec\chi)\in \mathbb{L}_2$ can be represented as $f(\vec{\chi})=\sum \limits_{j=1}^4
f_j(\vec{\chi}).$

\section{Decomposition of solutions}\label{sec:solutions}
In the present section we give exact decompositions of solutions of the wave equation based on the
formulas discussed above. The decomposition is associated with the boundary problem for the wave
equation in a half-plane $y\ge0$, $-\infty<x<\infty$:
\begin{eqnarray}
&u_{tt}=c^2(u_{xx}+u_{yy}) \label{eq:statement}\\
&u(ct, x, y)\left.\right|_{y=0}=f(ct, x), \label{eq:bound}\\
&u(ct, x, y)\left.\right|_{t \le 0} =0. \label{initial}
\end{eqnarray}
where $c$ is the speed, $t$ is  time, $x$ and $y$ are  spatial variables. The function $f(ct, x)$
in (\ref{eq:bound})  is
assumed to be square integrable:
\begin{equation}
\intd (ct) \intd x \, |f(ct, x)|^2 <\infty .
\end{equation}
The desired decomposition is expressed in terms of solutions running away from the boundary under
different angles. It  was first reported in \cite{Perel-2009}.

We define four subspaces of solutions $\mathfrak{H}_j,$ $j=1,\ldots,4$   of the wave equation
using the Fourier transform of the boundary data $\hat{f}(\vec\sigma)$. Any
function $u_{j}(\vec\chi,y ) \in \mathbb{L}_2$ belongs to $\mathfrak{H}_j$ if its Fourier
decomposition  reads
\begin{eqnarray}
&u_{j}(\vec\chi,y ) = \int\limits_{{\cal D}_{j}}  \rmd^2\sigma \,
 \hat{f}(\vec\sigma) \rme^{ \rmi (\vec\sigma, \vec\chi)}
  \rme^ {  \rmi\sqrt{k^2 - k_x^2} y}, \quad j =1,2,
 \label{eq:sol-1-2}
\\
&u_j(\vec\chi,y ) = \int\limits_{{\cal D}_{j}}  \rmd^2\sigma \,
 \hat{f}(\vec\sigma) \rme^{ \rmi (\vec\sigma, \vec\chi)} \rme^ {-\sqrt{k_x^2 - k^2} y},
 \quad j=3,4,  \label{eq:sol-3-4}
\end{eqnarray}
where $k = \omega/c$ is the wave number for any $\hat{f}(\vec\sigma)\in \mathbb{L}_2$. The  
functions $u_{j}(\vec\chi,y ), j=1,\ldots,4,$ are solutions
of the wave equation. If a function $\hat{f}(\vec\sigma)$ decreases at infinity not so slowly
to produce a classical solution, the functions $u_{j}(\vec\chi,y )$ satisfy the wave equation in 
the
distribution sense. The solutions
$u_{j}(\vec\chi,y ), j=1,2,$ are propagating, the solutions $u_{j}(\vec\chi,y ), j=3,4,$ are
exponentially decreasing and associated with details of the spatial behavior of  $f,$  which are
smaller than the wavelength. Note that the reconstruction formula for the Fourier transform yields
$\sum\limits_{j=1}^{4} u_j(\vec\chi,0) = f(\vec\chi).$

In what follows we will use the Fourier transforms of solutions with respect to $t$ and $x$ 
denoted by
$\hat{u}_j(\vec\sigma;y)$ :
\begin{eqnarray}
\hat{u}_j(\vec\sigma;y) & = \hat{f}(\vec\sigma) \rme^ {  \rmi\sqrt{k^2 - k_x^2} y},
j=1,2,\label{eq:Four-1-2}
\\
\hat{u}_j(\vec\sigma;y) & = \hat{f}(\vec\sigma) \rme^ {-\sqrt{k_x^2 - k^2} y},
j=3,4.\label{eq:Four-3-4}
\end{eqnarray}
The Fourier transforms of solutions with respect to $x$ and $y$ dependent on $t$ are denoted by
$\hat{u}_j(ct;k_x,k_y)$:
\begin{eqnarray}
\hat{u}_j(ct;k_x,k_y) & = \hat{f}(\vec\sigma) \rme^ { - \rmi \omega t}, j=1,2,
\quad \vec\sigma=(\sqrt{k_x^2 + k_y^2},k_x)^\trans.
\end{eqnarray}
The solutions ${u}_j(ct,x,y),$ $j=3,4,$ increase exponentially for negative $y$. The Fourier
transform of them with respect to $x$ and $y$ do not exist.

To develop a procedure of continuous wavelet analysis in the subspace $\mathfrak{H}_j,$ we should
choose a solution $\Psi_j \in \mathfrak{H}_j.$ We call it the mother solution. The notation of
the mother solution begins with the Greek uppercase letter and that distinguishes it from the 
mother
wavelet in $\mathbb{L}_2,$ which is denoted by the Greek lowercase letter.
The mother solution can be reduced to the mother wavelet if $y=0$. Therefore
$\psi_j(\vec\chi)=\Psi_j(\vec\chi,0)$ should satisfy the admissibility condition
(\ref{eq:admissibility_cond}).

Then we construct a family of solutions:
\begin{equation}
\Psi_{j\mu}(\vec\chi, y)=\frac{1}{a}\Psi_j\left(\Lambda_{-\phi} \frac{\vec{\chi} - \vec{b}}{a},\,
\frac{y}{a}\right).
\end{equation}
Each solution represents a wave packet considered in the stationary frame, which is a shifted   
and  scaled mother solution in the frame moving in the $x$ direction.
These solutions are transformed to  wavelets from the family (\ref{eq:wavelet_family}) on the line
$y=0$, i.e., $\Psi_{j\mu}(\vec\chi, 0)\equiv\psi_{j\mu}(\vec\chi),$  where $\psi_{j\mu}(\vec\chi)$
are wavelets.

  The decomposition of solutions of \eqref{eq:statement} in terms of mother wavelets  read
\begin{eqnarray}\label{eq:solution}
&u(\vec\chi, y)=\sum\limits_{j=1}^{4} u_j(\vec\chi, y), \quad
u_j(\vec \chi, y)=\frac{1}{{C_{j}}} \int \rmd\mu \,
F_j(\mu)\Psi_{j\mu}(\vec{\chi},y), \\
&F_j(\mu)=
\int\limits_{\mathbb{R}^2} \rmd^2 \vec{\chi} \, f(\vec\chi) \overline{\Psi_{j\mu}(\vec{\chi}, 0)}.
\end{eqnarray}
It is a solution because it is a decomposition of solutions. It is reduced to the reconstruction
formula for the boundary data $f$ if $y=0$ and therefore it satisfies the boundary conditions.
Each of solutions moves or decreases away from the boundary. We do not take into account
solutions that come from infinity and increase in the positive $y$ direction to satisfy initial 
conditions \eref{initial}.

 The contribution of each solution is determined by the coefficient $F_j,$  which is the wavelet
transform of the boundary data.

\section{Examples of mother solutions}\label{sec:wav_construct}

Mother solutions can be taken from a wide class. We give examples of mother solutions, which
behave as particles. As a starting point we take a solution that has the Fourier transform
localized near the point $k_x = 0,$ $k_y = \varkappa:$
\begin{equation}\label{eq:mother-gauss}
\hat\Psi(ct;k_x, k_y) =  \exp{\left( -\frac{\sigma_\parallel^2 (k_y-\varkappa)^2}{2} -
                                     \frac{\sigma_\perp^2 k_x^2}{2}\right)} \rme^{-\rmi \omega t},
\end{equation}
where $\omega = c |\vec{k}|.$
If $t=0$, then the solution in the coordinate domain reads
\begin{equation}\label{eq:psi1_t0_fourier}
\Psi(0, x, y) = \frac{1}{2\pi \sigma_\parallel \sigma_\perp}
 \exp{\left( {\rmi} \varkappa y  -
\frac{ y ^2}{2\sigma_\parallel^2} -
\frac{x^2 }{2\sigma_\perp^2} \right) }.
\end{equation}
If $t \ne 0$, the solution cannot be found explicitly.
But if $t$ is small enough, then $\Psi_{1}(ct,x,y) \approx \Psi_{1}(0,x,y-ct).$
It can be obtained by
expanding $k = \sqrt{k_x^2 + k_y^2}$ up to  linear terms: $\omega = \omega_0 + c (k_y - \varkappa)
+ \ldots$ and substituting  it into the phase:
\begin{equation}\nonumber
k_x x + k_y y - \omega t =  {k_x}x + k_y (y - c t) +\ldots.
\end{equation}
 On this way,  for small time we obtain a packet-like solution which is filled with oscillations of
 a spatial frequency $\varkappa$ and which has the Gaussian envelope. It moves along the $y$ axis 
 with speed $c$.

For large $t$ the solution should be found numerically. Exact solutions having similar local
behavior were found in \cite{Kiselev-Perel} and were named 'Gaussian wave packets'. Solutions and
their Fourier transforms  were given by  explicit formulas and were studied analytically. An
investigation of these solutions from the point of view of continuous wavelet analysis was 
performed in \cite{Per-Sid-JPA1}.

\begin{figure}[htbp]
\subfloat[Solutions in the moving frame]
{\includegraphics[width=0.49\textwidth]{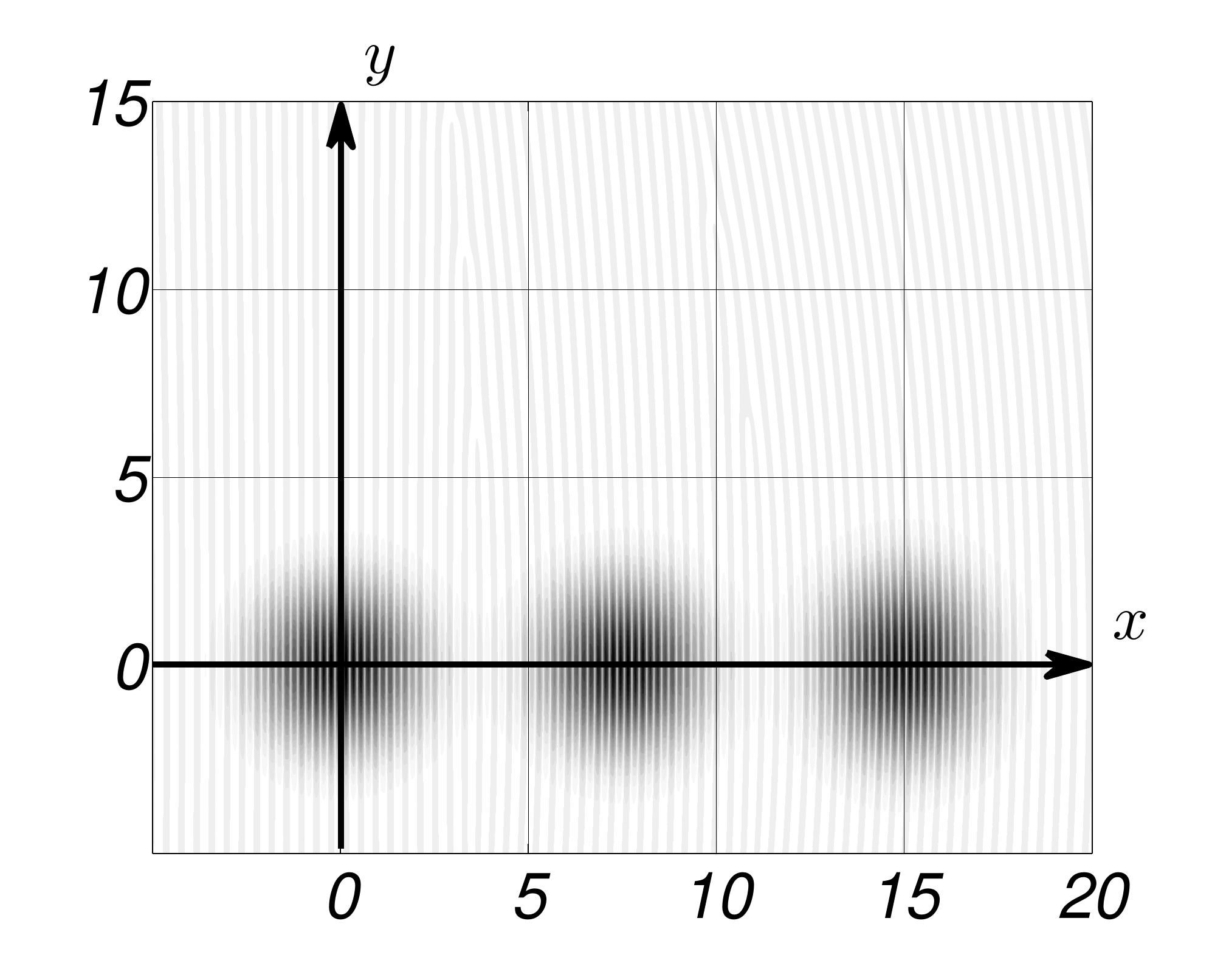}\label{fig:psi1:own}}
\subfloat[Same solutions in the stationary frame, $\phi=1$]
{\includegraphics[width=0.49\textwidth]{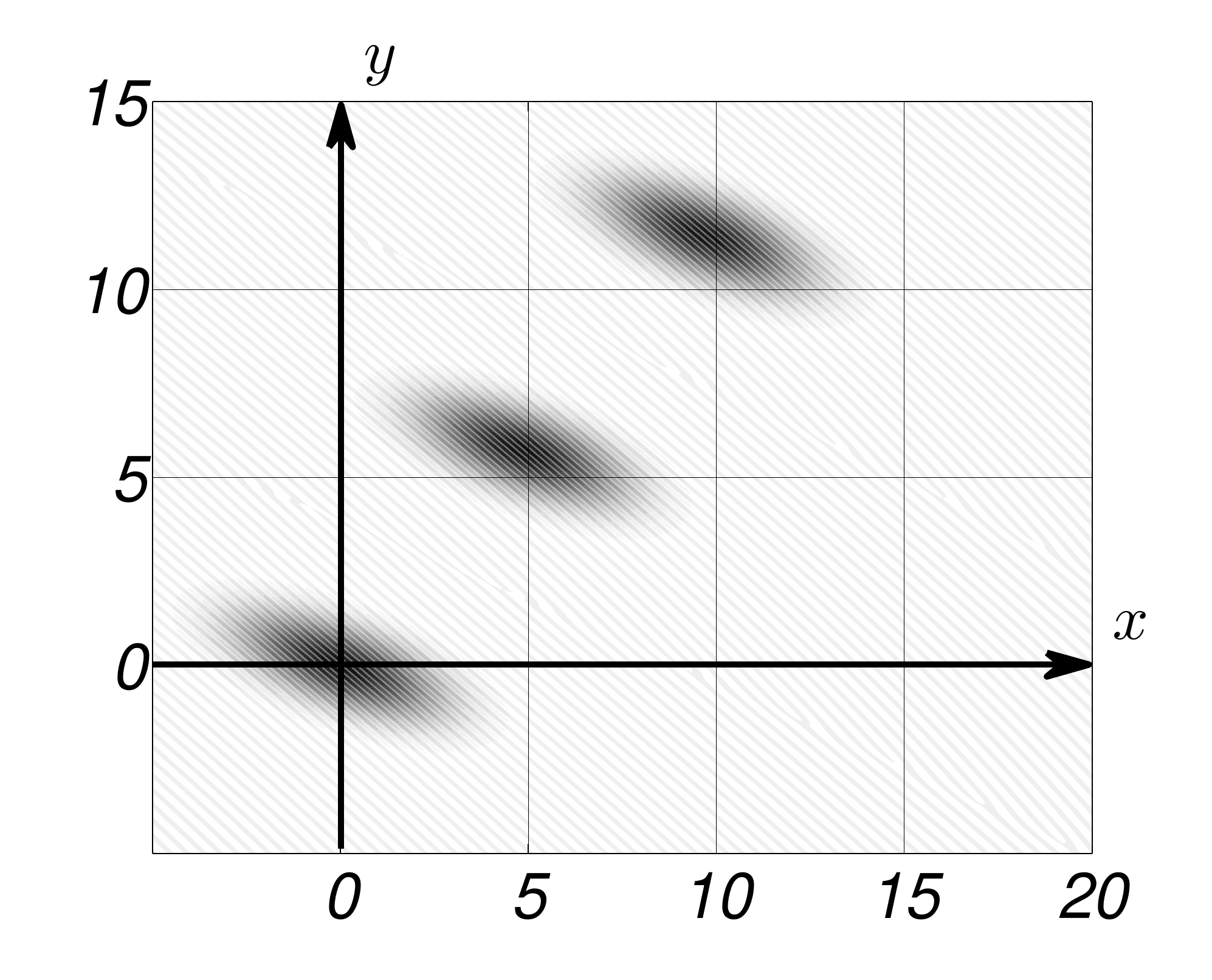}\label{fig:psi1:moving}}
\caption{Real part of mother wavelet $\Psi_1$ with parameters $\varkappa=16$,
$\sigma_\parallel=\sqrt{2}$, $\sigma_\perp=\sqrt{2}$ in successive time moments: $ct=0$, $ct=7.5$,
$ct=15$.
}
\label{fig:psi1}
\end{figure}

\begin{figure}[htbp]
\subfloat[Solutions in the moving frame]
{\includegraphics[width=0.49\textwidth]{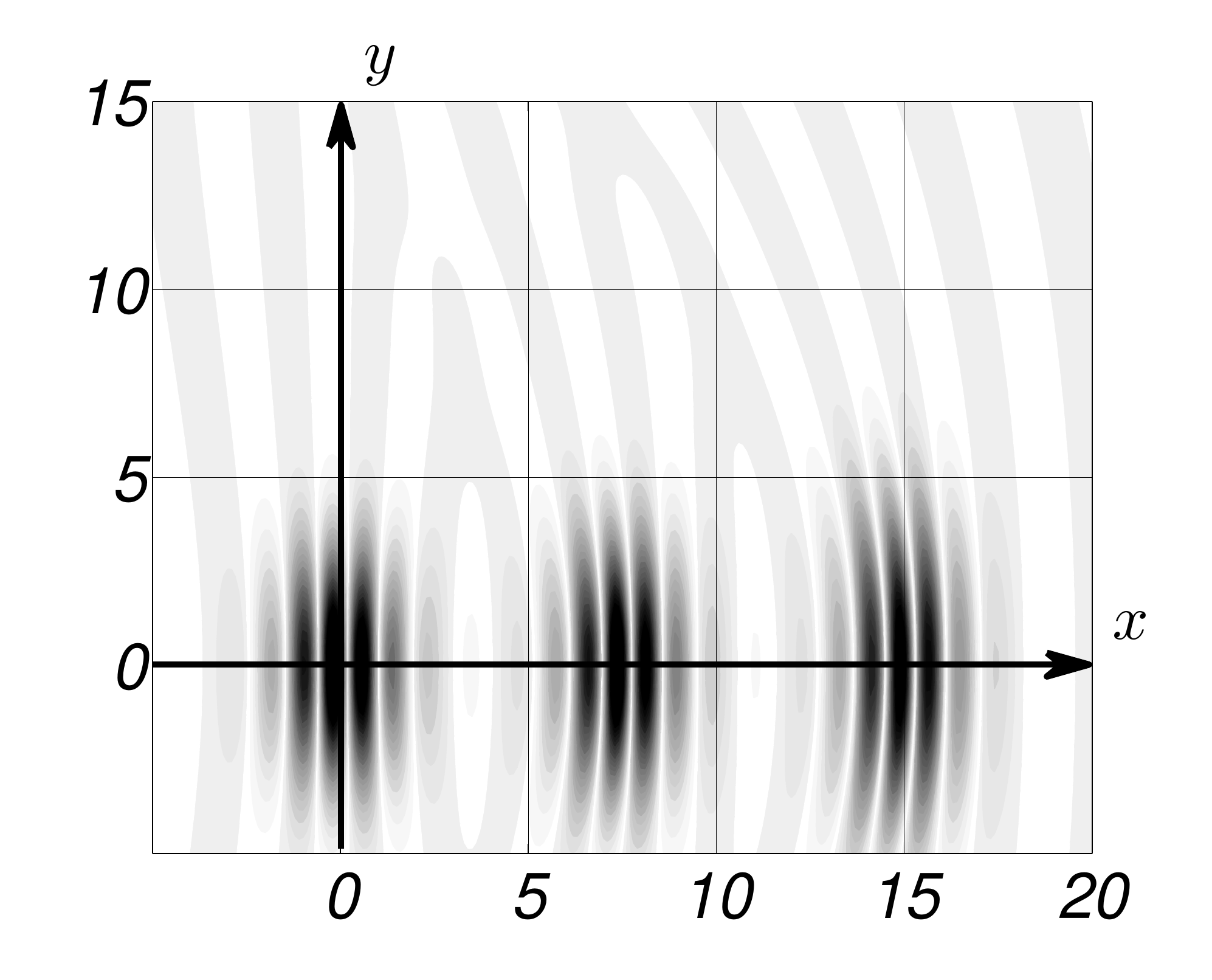}\label{fig:psi1k4:own}}
\subfloat[The same solutions in the stationary frame, $\phi=1$]
{\includegraphics[width=0.49\textwidth]{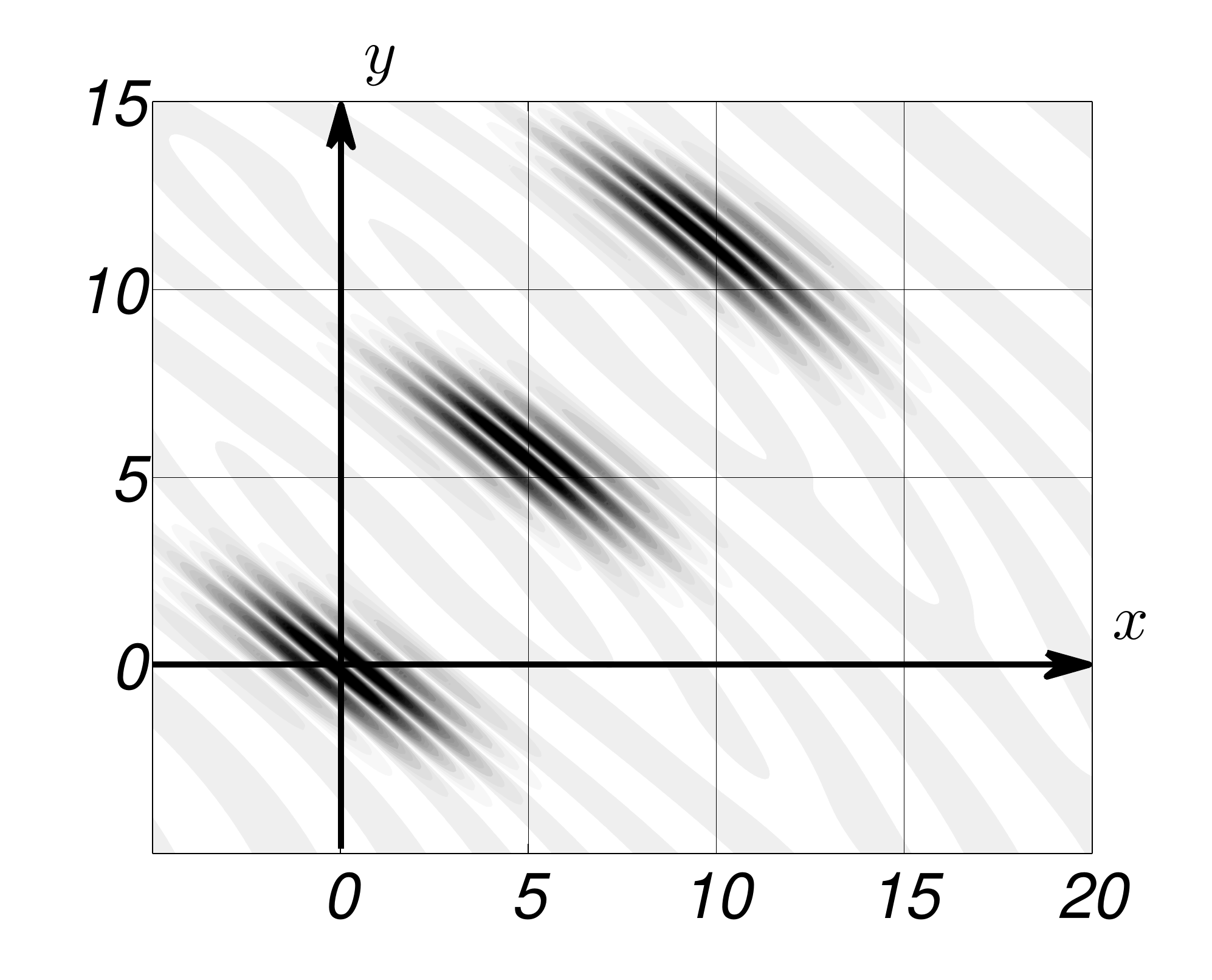}\label{fig:psi1k4:moving}}
\caption{Real part of the mother wavelet $\Psi^1$ with parameters $\varkappa=4$, 
$\sigma_\parallel=1$,
$\sigma_\perp=2$ in successive time moments: $ct=0$, $ct=7.5$,
$ct=15$.}
\label{fig:psi1k4}
\end{figure}

To be a mother solution in $\mathfrak{H}_1,$ this solution should contain Fourier components with
$\omega>0$ and $k_y>0$. The last condition is fulfilled only approximately if the central
frequency of the packet $\varkappa$ is larger than the width of the support of the Fourier
transform in $k_y$ direction $1/\sigma_{\parallel}$,
i.e. $\varkappa \sigma_{\parallel} =\sqrt {p}>> 1$. To obtain the mother solution, we assume that
formula \eqref{eq:mother-gauss} is valid only for
$k_y>0$ and $\hat\Psi(ct;kx, ky) = 0$ if $k_y \le 0.$  To make the Fourier transform smooth, we 
introduce  an additional factor  $\exp{(-\frac{1}{k_y})}, k_y = \sqrt{k^2
- k_x^2}.$  The mother solution $\Psi_{1}(ct,x,y)$ is of the form
\begin{eqnarray}\label{eq:psi1}
\fl
\Psi_{1}(ct,x,y)=\frac{1}{(2\pi)^2} \int \limits_{-\infty}^{\infty} \rmd k_x
                                               \int\limits_{0}^{\infty} \rmd k_y \,
\exp\left( -\frac{\sigma_\parallel^2 (k_y-\varkappa)^2}{2} -
                                     \frac{\sigma_\perp^2 k_x^2}{2}\right)
 \\ \nonumber
 \exp(- 1/k_y)\exp\left( \rmi (k_x x + k_y y -  kct) \right),
\end{eqnarray}
 $k=|\vec{k}|.$ Because of the factor $\exp{(- {1}/{k_y})}$, the solution satisfies the
 admissibility condition.
 We take a variable of integration $k=\sqrt{k_x^2 + k_y^2}$ instead of $k_y$ and write the Fourier
 transform of \eqref{eq:psi1} in the form of \eqref{eq:Four-1-2}. Then we obtain
\begin{eqnarray}
 \widehat\Psi_{1}(\vec{\sigma};y)= \frac{k}{k_y}
 \exp\left( -\frac{\sigma_\parallel^2(k_y-\varkappa)^2}{2} - \frac{\sigma_\perp^2 k_x^2}{2} -
 \frac{1}{k_y}\right)
\cdot \rme^{ \rmi  k_y y  },
\end{eqnarray}
$k_y = \sqrt{k^2 - k_x^2}$.
\Fref{fig:psi1:own}  shows this solution in successive time moments.
It represents a wave packet moving away from the boundary $y=0$.  It has a finite energy, because
it is localized in an exponential way. The same solution for other choice of parameters in the 
case $\varkappa \sigma_{\parallel} =\sqrt {p}=2 $ is shown in  \Fref{fig:psi1k4:own}.

\begin{figure}[htbp]
\subfloat[Solutions in the moving frame]
{\includegraphics[width=0.49\textwidth]{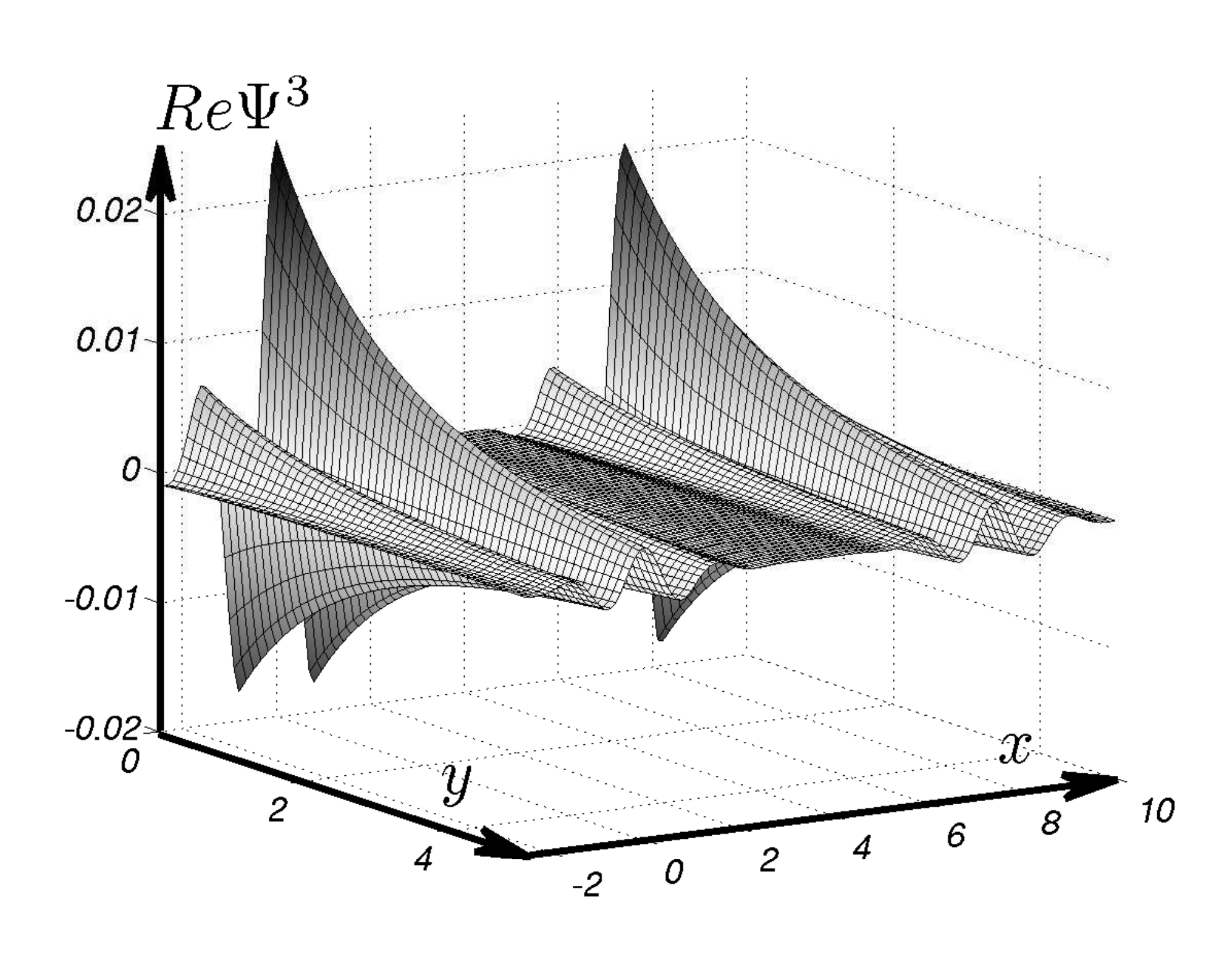}\label{fig:psi3k4:own}}
\subfloat[The same solutions in the stationary frame, $\phi=1$]
{\includegraphics[width=0.49\textwidth]{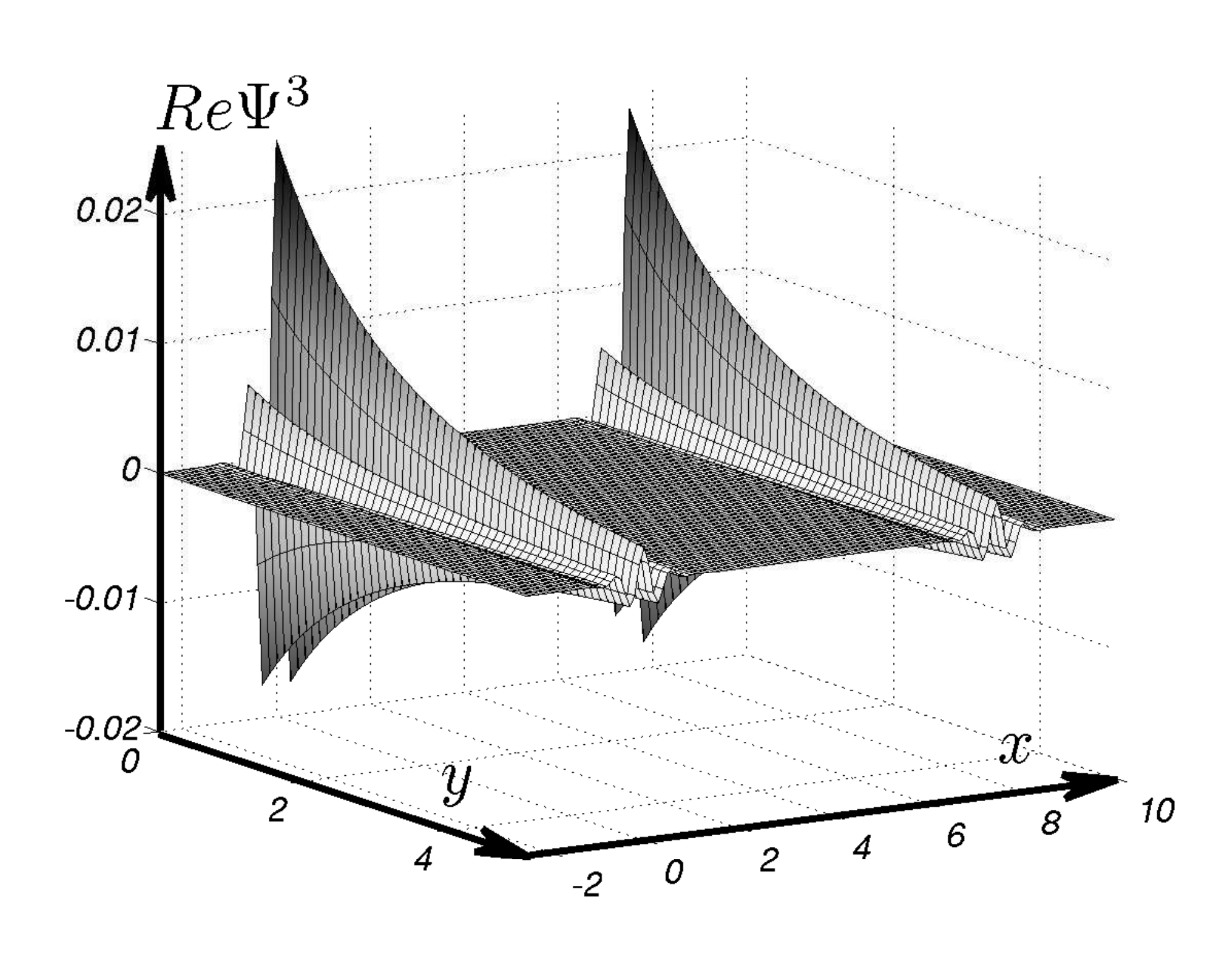}\label{fig:psi3k4:moving}}
\caption{Real part of the mother wavelet $\Psi_3$ with parameters $\varkappa=4$, 
$\sigma_\parallel=1$,
$\sigma_\perp=2$ in successive time moments: $ct=0$ and $ct=7.5$. }
\label{fig:psi3_3d}
\end{figure}

  The mother solution in $\mathfrak{H}_3$ is constructed in a similar way. We interchange $k$ and
 $k_x,$  put $iy$ instead of $y,$ and obtain
\begin{eqnarray}
\fl
 \hat\Psi_{3}(\vec{\sigma};y)= \frac{k_x}{\sqrt{k_x^2 - k^2}}
 \exp\left( -\frac{\sigma_\parallel^2
 (\sqrt{k_x^2-k^2}-\varkappa)^2}{2} - \frac{\sigma_\perp^2 k^2}{2} \right)
 \\ \nonumber
 \exp\left( -1/\sqrt{k_x^2-k^2}\right)\cdot \rme^{ -  \sqrt{k_x^2-k^2} y  }.
\end{eqnarray}
The solution $\Psi_{3}(ct,x,y)$ is a wave packet running along the boundary and decreasing
exponentially in the positive $y$ direction. The Fourier transform of the solution comprises
components with wavelengths in the $x$ direction, which are smaller than the wavelength in vacuum.
\Fref{fig:psi3k4:own} demonstrates the solution in successive time moments.
 The wavelet analysis of solutions from $\mathfrak{H}_3$ and $\mathfrak{H}_4$ might be
 useful for analyzing of surface waves, but a more detail study of the problem goes beyond the 
 scope of this paper.

The mother solutions in $\mathfrak{H}_2$ and $\mathfrak{H}_4$ can be obtained as follows:
$\Psi_{2}(ct,x,y)=\Psi_{1}(-ct,x,y),$ $\Psi_{4}(ct,x,y)=\Psi_{3}(ct,-x,y).$

Now we discuss analytically the properties of the family of solutions obtained by means of  Lorentz
transformations.
The mother solution $\Psi_{1}(ct,x,y)$ represents a wave packet moving in the positive $y$
direction and localized near the point $x=0$, $y=ct$.
In the Fourier domain, it is localized near the point $\sigma_0=(\varkappa, 0),$
$\omega_0 = c \varkappa.$
Now consider  $\Psi_{1}(ct',x',y),$ where $x',y,t'$ are coordinates in the frame that moves with 
speed $v$ and  rapidity $\phi.$ The coordinates in the moving frame are connected with the 
coordinates in the stationary frame by transformations (\ref{eq:lorentz}). Assuming that $t'$ is 
small enough, we obtain
$\Psi_{1}(ct',x',y) \approx \Psi_{1}(0,x',y-ct') =
\Psi_{1}(0,\cosh\phi(x - ct \tanh\phi),y + \sinh\phi x - \cosh\phi ct).$  The position of the
center of the packet $x_0, y_0$ in the stationary frame is  $x_0 = ct \tanh \phi,$ $y_0 = ct/\cosh
\phi.$   The packet \eqref{eq:psi1} found in the moving frame is numerically small outside an
ellipse
\begin{equation}
\frac{ (y - c t')^2}{2\sigma_\parallel^2} +
\frac{x'^2 }{2\sigma_\perp^2} \le 1
\end{equation}
For the sake of simplicity we assume that it is a circle $\sigma \equiv \sigma_\parallel =
\sigma_\perp$:  
\begin{equation} \label{eq:circ}
 (x')^2 +  (y - ct')^2 \le 2\sigma^2.
\end{equation}
An example of such a solution is shown in \Fref{fig:psi1:own}.
Taking into account (\ref{eq:lorentz}) we find  the domain in the stationary frame corresponding
to the circle (\ref{eq:circ}) in the moving frame.  It is an ellipse:
\begin{eqnarray}\label{eq:circ-form}
&\cosh^2\phi (x - x_0)^2 + ( (y-y_0) - \sinh\phi (x-x_0) )^2 \le  2\sigma^2.
\end{eqnarray}
This ellipse in the main axes reads
\begin{equation}
\lambda_1 \tilde{x}^2 + \lambda_2 \tilde{y}^2 \le  2\sigma^2.
\end{equation}
 where \begin{equation}
\lambda_{1,2}= \rme^{\pm \phi} \cosh\phi.
\end{equation}
The directions of the main axes are
\begin{equation}
\vec{e_{1,2}}=\left(\begin{array}{c}
                   \pm \rme^{\mp \phi}
                   \\
                    1
              \end{array} \right).
\end{equation}
The new coordinates  $\tilde{x}$, $\tilde{y}$  are connected with the old coordinates $x-x_0,$ 
$y-y_0$
by the relation
\begin{equation}
\binom{\tilde{x}}{\tilde{y}} = U^\trans \binom{x-x_0}{y-y_0}, \quad U =
\left(\frac{\vec{e}_1}{|\vec{e}_1|}, \frac{\vec{e}_2}{|\vec{e}_2|}\right),
\end{equation}
where the columns of the matrix $U$ are vectors $\vec{e_j}, j=1,2.$

\begin{figure}[htbp]
\centering
\includegraphics[keepaspectratio, width=0.75\textwidth]
{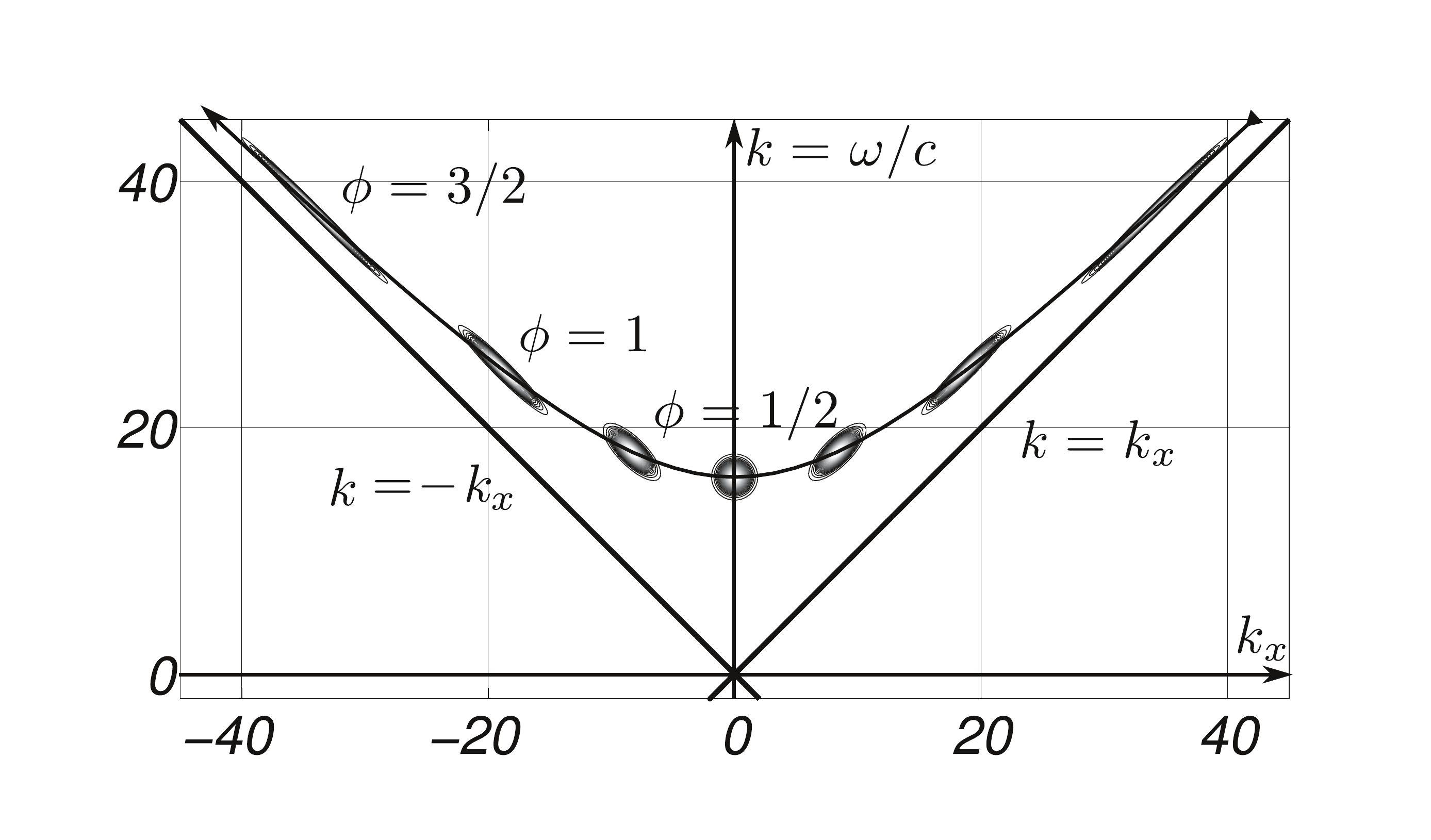}
\caption{The support of the wavelet in the Fourier domain for various $\phi$ }
\label{fig:lorentz}
\end{figure}

The axes of the ellipse in the stationary frame are not orthogonal to the direction of propagation
and to wave fronts (see \Fref{fig:psi1}).

Similar considerations are valid in the Fourier domain. The support of the Fourier transform
of the solution is distorted under the action of the Lorentz transform. Examples of calculations
for various $\phi$ are given in  \Fref{fig:lorentz}; a discussion of \Fref{fig:lorentz} see below.

\section{Numerical calculations of coefficients in the wavelet decompositions}\label{sec:examples}

A decomposition of solutions is given by a fourfold  integral \eref{eq:solution}. Here we show that
calculations of such an integral can be efficient for some boundary data.  For realistic signals 
the
domain of integration may be sparse. The domain of integration is determined by the numerical
support of the coefficient in decomposition \eref{eq:solution}, which  is the wavelet transform of
boundary data.
The wavelet transform is non-negligible if the supports of  a signal (the domain where the signal 
is
non-zero) and of a wavelet from the family  \eref{eq:wavelet_family} have an
intersection.
The supports of Fourier transforms of a signal and a wavelet should also have an intersection.

The support in the Fourier domain is governed by a scale and a rapidity. Figure \Fref{fig:lorentz}
demonstrates transformations of the support of the wavelet in the Fourier domain for different
values of the rapidity $\phi$. The support for $\phi=0$ is a circle. Upon the Lorentz
transformations, the support takes the form of an ellipse and its center moves along the hyperbola.
Here there is an analogy with the transformations of the support in a spatial domain. A change in 
the scaling parameter $a$ will lead to a change in the hyperbola, where the centers of supports are
located.

Shifts in the spatial domain $b$ rule oscillations of an integrand in \eref{eq:Poincare_CWT}. An 
increase in
oscillations  results in a decrease in the wavelet transform and vice versa.
If the support of a wavelet for some parameters matches the support of a signal and $\vec{b}$ is
taken in such a way that oscillations are minimized, then the wavelet transform has a maximum.

\begin{figure}[htbp]
\subfloat[The function $f(ct,x)$]
{\includegraphics[keepaspectratio, width=.49\textwidth]
{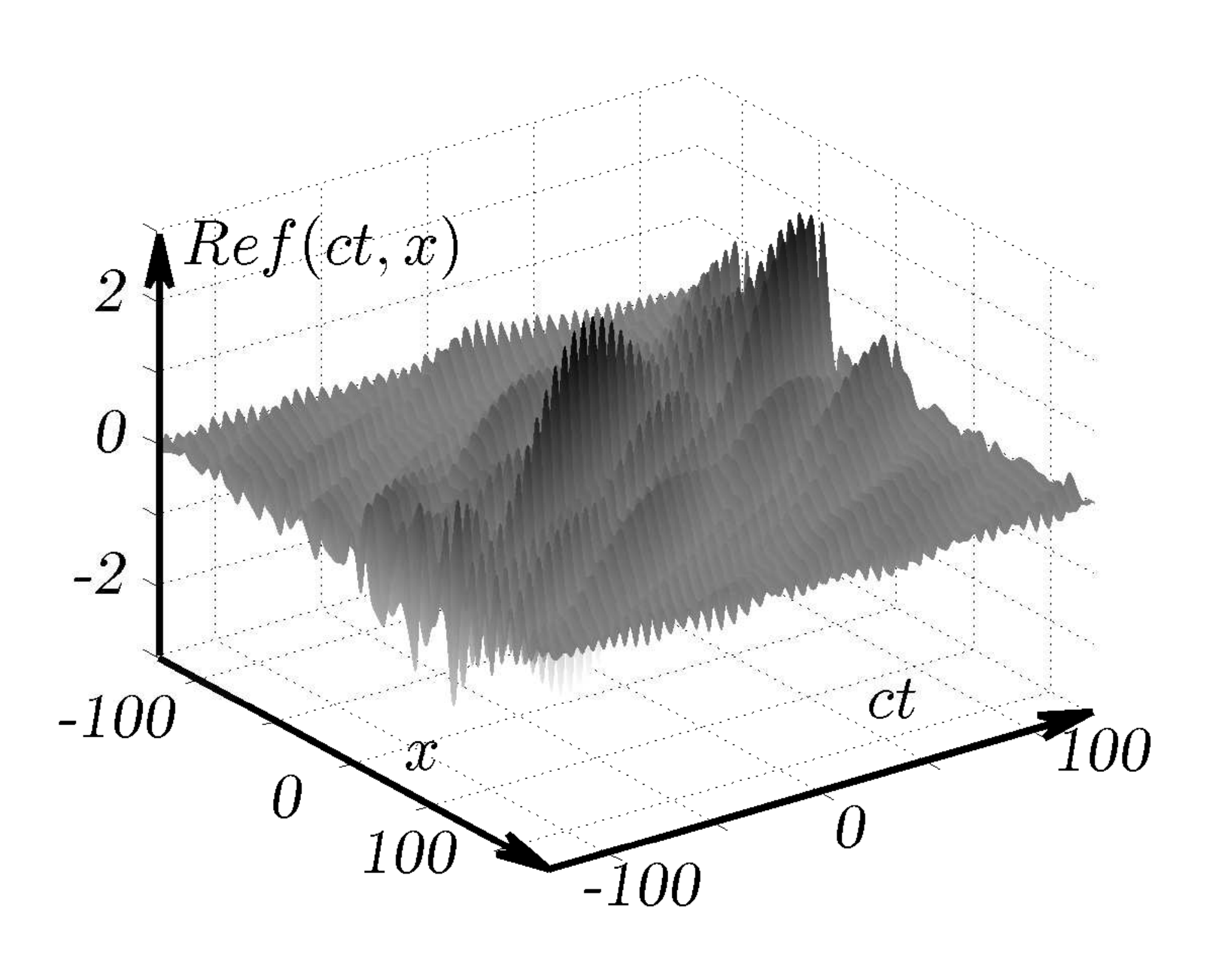}\label{fig:example:signal}}
\subfloat[The scale-rapidity diagram  $S_1(a,\phi)$]
{\includegraphics[keepaspectratio, width=.49\textwidth]
{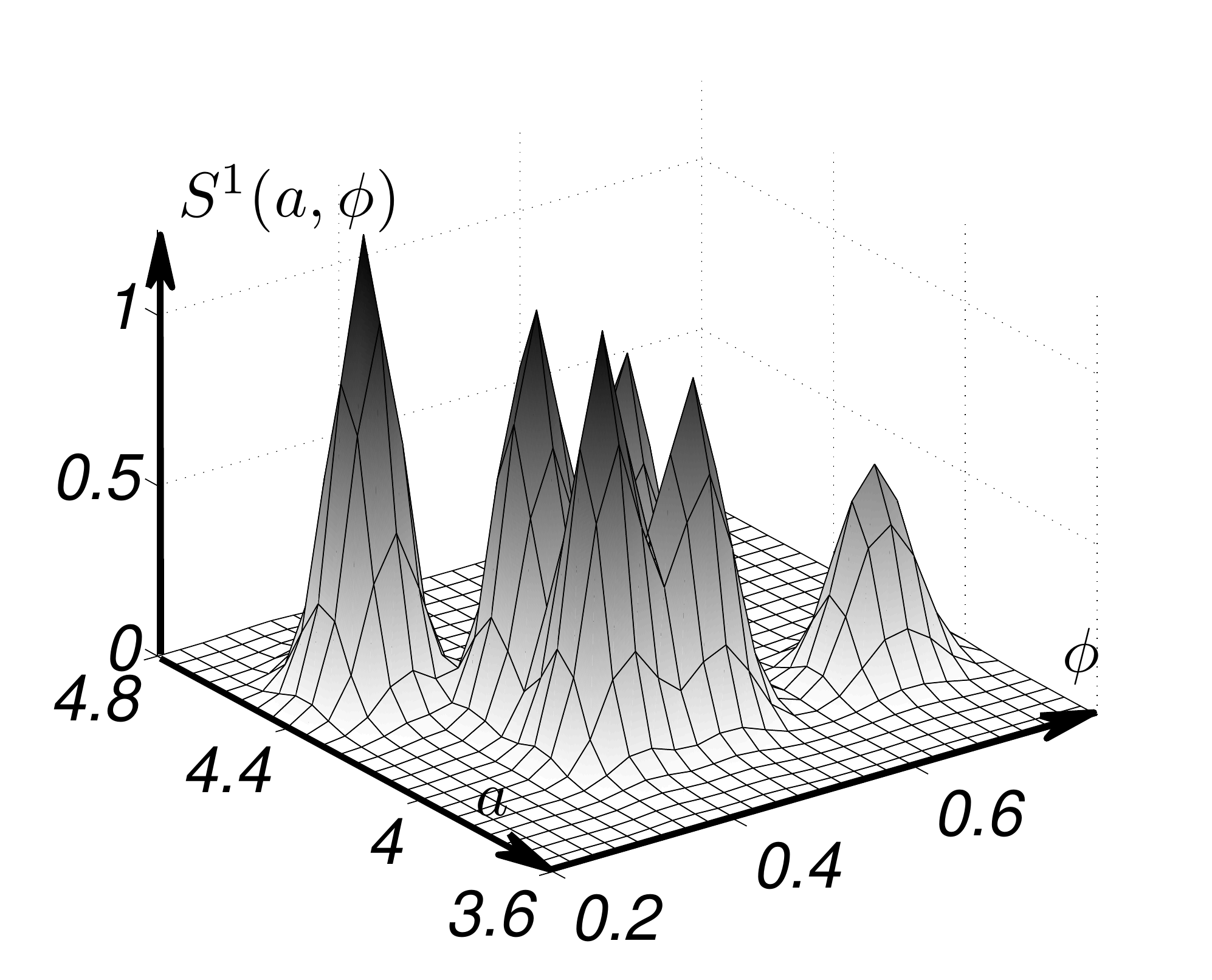}\label{fig:example:sc_gamma} }
\caption{An example of the scale-rapidity diagram.}
\label{fig:example}
\end{figure}

Now consider  an example. We assume that a function $f$ represents a field that is generated in the
plane $y=0$  by  six groups of monochromatic  point sources, which move in the $x$
direction with different speeds in plane $y=-5000$. In every group, the speeds are distributed
with respect to the Gaussian law with $\sigma = 0.01 c,$ $\sigma$ is a dispersion of the
distribution, $c$ is the speed in the wave equation, $c=1$. The sources have
different frequencies. The sources in the $j$th group,  $j=1,\ldots, 6,$ are characterized by  the 
frequency
 $\omega_j$ and  the rapidity $\phi_j,$ which corresponds to the mean speed. We choose
the parameters as follows: $\omega_1=1, \phi_1=0.4$;
$\omega_2=1, \phi_2=0.7$; $\omega_3=1, \phi_3=0.5$; $\omega_4=0.9, \phi_4=0.3$;
$\omega_5=0.95, \phi_5=0.5$; $\omega_6=0.95, \phi_6=0.4$.
The boundary function $f$ as a function of $x$ and $t$ is shown in \Fref{fig:example:signal}.
  We show that, analyzing the field in
the plane $y=0$ by
means of the wavelet transform, we can
determine the parameters of the sources. The sources are far from the observation plane, and we do 
not
calculate the wavelet transforms in domains ${ \cal{D}}_{3,4}$, because the contribution of the 
components
propagating along the surface are negligible.

We show that the wavelet transform enables us to
find the frequencies and speeds of sources. We are not interested in the determination of the 
position
of sources; for this reason  we introduce  a function
$S(a,\phi)$, which reads
\begin{equation}
S_1(a,\phi) = \frac{1}{{a^3}}{\int\limits_{\R^2} d^2\vec b \
|F_1(\mu)|^2}, \quad \mu=\{ \vec{b}, a, \phi \}.
\end{equation}
 Such a function, corresponding to  the field in \Fref{fig:example:signal},
 as a function of the scaling $a$ and the rapidity $\phi$ is given in \Fref{fig:example:sc_gamma}. 
 The
 scale-rapidity
 diagram shown is calculated with the following wavelet parameters: $\varkappa=4,
 \sigma_\parallel=2\sqrt{55}\approx 15, \sigma_\perp=8$,
 the calculation mesh is
 from $-128$ to $128$ with step $0.5$ in both $x$ and $ct$.
We see six maxima in this figure, which correspond to each group of  sources. The position of the
maximum $a,\phi$ contains
information about the frequency $\omega = 1/a$ and the average rapidity $\phi$ of sources.
(We recall that the speed $v$ is connected with $\phi$ by the relation $v/c = \tanh \phi.$)
The domain of integration in the decomposition formula must contain  values of the parameters $a$, 
$\phi$ such that $S$ is not small.  If we are interested only in the investigation of sources of
certain frequencies and rapidities, we need to take the corresponding domain of integration.

A more complicated situation where the sources are masked by noise are considered by us in
\cite{Gorodnitskiy-Perel-2011}.
It is shown therein that the contribution of a noise can be eliminated, because it is located at
small scales on the scale-rapidity diagram. The parameters of sources are found. 

\ack
E. Gorodnitskiy indebted to Dr. Ru-Shan Wu  for his
hospitality in University of California, Santa-Cruz which made possible a completion of the text 
and figures preparation.
\appendix
\section{The reconstruction formula}\label{app:rec}
We prove the Plancherel equality \eqref{eq:Plancherel_eq} for the affine Poincar\'e wavelet 
transform
\eqref{eq:Poincare_CWT} to make the paper self-contained. For the sake of brevity, we assume
that $f\equiv g$. With account of  \eqref{eq:dmu} the Plancherel equality reads
\begin{equation}\label{eq:app-Plan}
{\intd \phi}
{\int \limits_{0}^{\infty} \frac{\rmd a}{a^3}}
{\int \limits_{\mathbb{R}^2} \rmd^2 \vec{ b} } \
\left| F_j(\vec{ b}, a, \phi) \right|^{2}=
C_j \intdd \vec{\chi} \left|f_j(\vec{\chi}) \right|^{2},
\end{equation}
where $C_j$  is defined by relation \eqref{eq:admissibility_cond}.
\\
Applying the Plancherel equality \eqref{eq:Plancherel_fourier} to the inner integral with respect
to $\vec{b}$ and taking into account the definition of $F_j,$ we obtain
\begin{equation}\label{eq:first-step}
{\int \limits_{\mathbb{R}^2} \rmd^2 \vec{ b} } \
\left| F_j(\vec{ b}, a, \phi) \right|^{2}=\frac{a^2} {(2 \pi)^2} \intdd{\sigma} \left|
\hat{f}(\sigma)\right|^2 \,\left|\hat{\psi}_j(a \Lambda_{-\phi}
\vec{\sigma}) \right|^2.
\end{equation}
Here we use the fact that $F_j$ is a convolution of two functions $f(\vec\chi)$ and
$\overline{\psi_{j\mu}}(-\vec\chi)$. Then it is used that the Fourier transform of
$\overline{\psi_{j\mu}}(-\vec\chi)$ is  $\overline{\widehat{\psi_{j\mu}}(\vec\sigma)},$ which is
calculated by the formula \eqref{eq:wav-Fourier}.

Substituting \eref{eq:first-step} in the left-hand side of \eref{eq:app-Plan} and interchanging
the order of integration, we obtain
\begin{equation}\label{eq:second-step}
\fl {\intd \phi}
{\int \limits_{0}^{\infty} \frac{\rmd a}{a^3}}
{\int \limits_{\mathbb{R}^2} \rmd^2 \vec{ b} } \
\left| F_j(\vec{ b}, a, \phi) \right|^{2}
=\frac{1} {(2 \pi)^2} \intdd{\vec\sigma} \left| \hat{f}(\vec\sigma)\right|^2 \intd{\phi}
\int\limits_{0}^{\infty} \frac{da}{a} \,
\left|\hat{\psi}_j(a \Lambda_{-\phi} \vec{\sigma}) \right|^2
\end{equation}
The next step is an analysis of the inner integral on the right-hand side of
\eref{eq:second-step}. For the sake of definiteness, we assume that  $j=1$.
 We transform the variable of integration in the following way:
 \begin{equation}\label{eq:var-sig}
 \fl \vec\sigma' = a \Lambda_{-\phi} \sigma = a \binom{k \cosh \phi+k_x \sinh \phi }{k \sinh \phi
 + k_x \cosh \phi} = a \binom{ \rho \cosh (\phi+\phi_0) } {\rho \sinh (\phi + \phi_0)}
 = \rho' \binom{\cosh\phi'}{\sinh\phi'}.
 \end{equation}
Here we use the notation: $k=\rho \cosh \phi_0$, $k_x=\rho \sinh \phi_0$, $\rho>0$.
 Instead of $a$ and $\phi$ we take new variables  $\phi'=\phi+ \phi_0$ and $\rho'=a \rho$, $\phi'
 \in (-\infty, +\infty),$ $\rho' \in (0, \infty).$  Then the vector variable $\vec\sigma'$ is
 chosen instead of $\phi',$ $\rho'.$ It takes values in ${\cal D}_1.$ The inner integral for any
 fixed $\vec\sigma$ is transformed as follows:
\begin{eqnarray}
 \fl & \intd{\phi} \int\limits_{0}^{\infty} \frac{\rmd a}{a} \,\left|\hat{\psi}_j(a \Lambda_{-\phi}
\vec{\sigma}) \right|^2=
\intd{\phi'} \int\limits_{0}^{\infty} \frac{\rho' \rmd \rho'}{\rho'^2}
\bigl|\hat{\psi}_j(\vec\sigma')\bigr|^2 \\
& =
\intdd{\vec\sigma'}
\frac {\bigl|\hat{\psi}_j( \vec{\sigma'})\bigr|^2}{k'^2-k'^2_x} \equiv C_j, \quad \vec\sigma' =
\binom{k'}{ k'_x}.
\end{eqnarray}

Formula \eqref{eq:Plancherel_eq} has a generalization, in which
 $f, g$ are any functions from $\mathbb{L}_2$ and the wavelet transforms of the functions $f$
and $g$ denoted by $F_j$ and $G_j,$ respectively, are calculated
with different mother wavelets, say $\zeta$ and $\psi,$ respectively.
In this case, formula \eref{eq:Plancherel_eq} is valid, but $C_j$ is calculated as follows:
\begin{equation}\label{eq:app-admissibility_cond}
C_j= \int\limits_{{\cal{D}}_j} \rmd^2 \vec{\sigma} \,
\frac{\overline{\hat{\zeta}_j(\vec\sigma)}\hat{\psi}_j(\vec\sigma)}{|(\omega/c)^2 - k^2_x|} <
\infty. \,
\end{equation}

The reconstruction formula \eref{eq:poinc-WT_reconst} may be obtained from \eref{eq:Plancherel_eq}
formally if we choose as a function $g$ any sequence of functions that tends to the Dirac
$\delta$ function. To pass to the limit under the sign of the integral in \eref{eq:Plancherel_eq},
we must impose additional restrictions on the functions $f$ and $g.$

\section{Ellipse transformations}\label{app:ellipse}
 Assume that the packet \eqref{eq:psi1} is  negligible outside the ellipse, i.e. if $\alpha^2 
(x')^2 + \beta^2 (y - ct')^2 \ge 1$ in the moving frame. In the stationary frame this ellipse is 
distorted as follows:
\begin{eqnarray}\label{eq:quadr-form}
&\alpha^2 \cosh^2\phi (x - x_0)^2 + \beta^2 ( (y-y_0) - \sinh\phi (x-x_0) )^2 \ge 1
\end{eqnarray}
The solution is located in the ellipse which reads in the main axes
\begin{equation}
\lambda_1 \tilde{x}^2 + \lambda_2 \tilde{y}^2 \le 1.
\end{equation}
 where $\lambda_j, j=1,2$ are eigenvalues   of the matrix of the quadratic form in the left-hand 
 side of the inequality \eqref{eq:quadr-form} which are as follows
\begin{equation}
\lambda_{1,2}=\frac{(\alpha ^2+\beta ^2) \cosh^2\phi}{2} \pm
  \frac{1}{2}\sqrt{(\alpha ^2+\beta^2)^2 \cosh^4\phi - 4 \alpha ^2 \beta ^2 \cosh^2\phi}).
\end{equation}
The directions of main axes read
\begin{equation}
\fl
\vec{e_{1,2}}=\left( \begin{array}{c}
                   			\beta^2 -  \alpha^2 \cosh^2\phi -\beta^2\sinh^2\phi \pm
                   			 \sqrt{(\alpha^2 + \beta^2)^2\cosh^4\phi - 4\alpha^2\beta^2\cosh^2\phi}
                   			\\
                   			 2\beta^2 \sinh\phi
                   	 \end{array}
              \right)     	 
\end{equation}
The new coordinates  $\tilde{x}$, $\tilde{y}$  are connected with old coordinates $x-x_0,$ $y-y_0$ 
by the relation
\begin{equation}
\binom{\tilde{x}}{\tilde{y}} = U^\trans \binom{x-x_0}{y-y_0}, \quad U = 
\left(\frac{\vec{e}_1}{|\vec{e}_1|}, \frac{\vec{e}_2}{|\vec{e}_2|}\right)
\end{equation}
where the columns of the matrix $U$ are vectors $\vec{e_j}, j=1,2.$


\end{document}